\documentclass[aps,prl,showpacs,amsmath,showkeys]{revtex4}
\usepackage{graphicx}

\newcommand{\ba}{\begin{eqnarray}}
\newcommand{\ea}{\end{eqnarray}}
\newcommand{\ban}{\begin{eqnarray*}}
\newcommand{\ean}{\end{eqnarray*}}
\newcommand{\bsub}{\begin{subequations}}
\newcommand{\esub}{\end{subequations}}

\begin{document}

\title{From Exact to Partial Dynamical Symmetries:\\ 
Lessons From the Interacting Boson Model}

\author{A. Leviatan}
\affiliation{
Racah Institute of Physics, The Hebrew University, 
Jerusalem 91904, Israel}

\begin{abstract}
We exploit the rich algebraic structure of the interacting boson model 
to explain the notion of partial dynamical symmetry (PDS), and present 
a procedure for constructing Hamiltonians with this property. 
We demonstrate the relevance of PDS 
to various topics in nuclear spectroscopy, 
including $K$-band splitting, odd-even staggering in the $\gamma$-band 
and anharmonicity of excited vibrational bands.
Special emphasis in this construction is paid to the role of 
higher-order terms.
\end{abstract}

\maketitle


\section{\textbf{Introduction}}

The concept of dynamical symmetry (DS)
is now widely accepted to be of central importance
in our understanding of many-body systems.
In particular, it had a major impact
on developments in nuclear~\cite{Iachello87}, 
molecular~\cite{Iachello95} and hadronic physics~\cite{BIL94}, 
pioneered by F.~Iachello and his colleagues. 
Its basic paradigm is to write the Hamiltonian
of the system under consideration
in terms of Casimir operators
of a set of nested algebras.
Its hallmarks are
(i)~solvability of the complete spectrum, 
(ii)~existence of exact quantum numbers for all eigenstates, and 
(iii)~pre-determined symmetry-based structure of the eigenfunctions,
independent of the Hamiltonian's parameters.

The merits of a DS are self-evident. However, in most applications to
realistic systems, the predictions of an exact DS are rarely fulfilled
and one is compelled to break it.
More often one finds that the assumed symmetry is
not obeyed uniformly, {\it i.e.}, is fulfilled by only some states 
but not by others. The need to address such situations has led to 
the introduction of
partial dynamical symmetries (PDSs)~\cite{Lev11}.
The essential idea is to relax the stringent conditions
of {\em complete} solvability
so that the properties (i)--(iii)
are only partially satisfied.

An exact symmetry occurs when the Hamiltonian of the system commutes 
with all the generators of the symmetry group $G$. 
In this case, all states have good symmetry and are labeled by the 
irreducible representations (irreps) of $G$. 
The Hamiltonian admits a block structure so that 
inequivalent irreps do not mix and all eigenstates 
in the same irrep are degenerate. 
In making the transition from an exact to a 
dynamical symmetry, states which 
are degenerate in the former scheme are split but not mixed 
in the latter, in accord with the reduction 
$G \supset G' \supset G''\supset...$, 
and the block structure of the Hamiltonian is retained.
Proceeding further to partial dynamical symmetry, some blocks or selected 
states in a block remain pure, while other states mix and lose the 
symmetry character. 
The hierarchy of broken symmetries and the corresponding Hamiltonian 
matrices are depicted schematically in Fig.~1.
\begin{figure}[t]
\includegraphics[width=\textwidth]{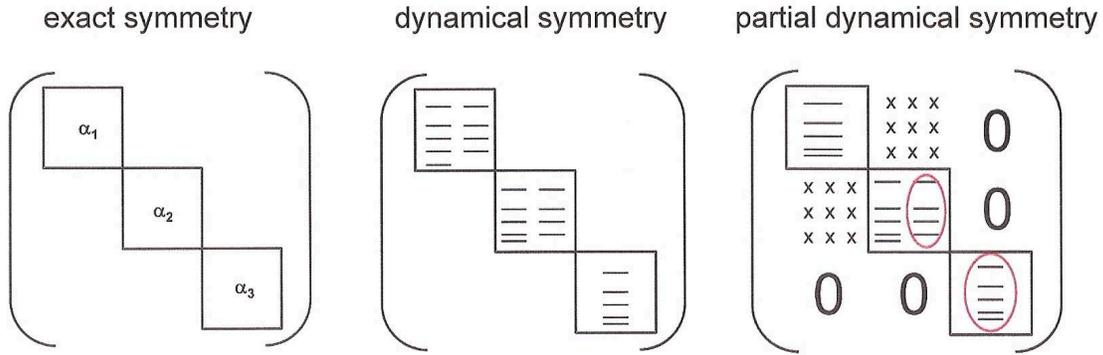}
\caption{The Hamiltonian structure in an exact, dynamical (DS) and partial 
dynamical symmetry (PDS). In an exact symmetry,  all states in a given 
irrep, $\alpha_i$, are degenerate. The DS exhibits splitting but no mixing. 
In a PDS,  only selected states (marked by an oval) remain solvable 
with good symmetry.}
\end{figure}

\section{\textbf{Identifying Hamiltonians with PDS}}

The algorithm for constructing Hamiltonians with PDS ~\cite{AL92,RamLevVan09}
starts from the chain of nested algebras
\begin{equation}
\begin{array}{ccccccc}
G_{\rm dyn}&\supset&G&\supset&\cdots&\supset&G_{\rm sym}\\
\downarrow&&\downarrow&&&&\downarrow\\[0mm]
[h]&&\langle\Sigma\rangle&&&&\Lambda
\end{array} ~,
\label{chain}
\end{equation}
where, below each algebra,
its associated labels of irreps are given. Eq.~(\ref{chain}) implies 
that $G_{\rm dyn}$ is the dynamical (spectrum generating) 
algebra of the 
system such that operators of all physical observables 
can be written in terms of its generators; 
a single irrep of $G_{\rm dyn}$
contains all states of relevance in the problem.
In contrast, $G_{\rm sym}$ is the symmetry algebra
and a single of its irreps contains states that are degenerate in energy. 
For $N$ identical particles the representation 
$[h]$ of the dynamical algebra $G_{\rm dyn}$ 
is either symmetric $[N]$ (bosons)
or antisymmetric $[1^N]$ (fermions)
and will be denoted, in both cases, as $[h_N]$. 
The occurrence of a DS of the type~(\ref{chain})
signifies that the Hamiltonian is written in terms of the Casimir 
operators of the algebras in the chain, 
$\hat{H}_{\rm DS} = \sum_{G} a_{G}\,\hat{C}(G)$, 
and the eigenstates can be labeled as
$\vert [h_N]\langle\Sigma\rangle\dots\Lambda\rangle$;
additional labels (indicated by $\dots$)
are suppressed in the following.
The eigenvalues of the Casimir operators in these basis states 
determine the eigenenergies $E_{\rm DS}([h_N]\langle\Sigma\rangle\Lambda)$ 
of $\hat{H}_{\rm DS}$. Likewise, operators can be classified
according to their tensor character under~(\ref{chain})
as $\hat T_{[h_n]\langle\sigma\rangle\lambda}$.

Of specific interest in the construction of a PDS
associated with the reduction~(\ref{chain}),
are the $n$-particle annihilation operators $\hat T$ 
which satisfy the property
\begin{equation}
\hat T_{[h_n]\langle\sigma\rangle\lambda}
\vert [h_N]\langle\Sigma_0\rangle\Lambda\rangle=0 ~,
\label{anni}
\end{equation}
for all possible values of $\Lambda$
contained in a given irrep~$\langle\Sigma_0\rangle$ of $G$. 
Any $n$-body, 
number-conserving normal-ordered interaction
written in terms of these annihilation operators 
and their Hermitian conjugates (which transform as the
corresponding conjugate irreps) 
can be added to the Hamiltonian with a DS~(\ref{chain}),
$\hat{H}_{\rm PDS} = \hat{H}_{\rm DS} + \sum_{\alpha,\beta} 
A_{\alpha\beta}\, \hat{T}^{\dag}_{\alpha}\hat{T}_{\beta}$, 
while still preserving the solvability
of states with $\langle\Sigma\rangle=\langle\Sigma_0\rangle$. 
The annihilation condition~(\ref{anni}) is satisfied 
if none of the $G$ irreps $\langle\Sigma\rangle$
contained in the $G_{\rm dyn}$ irrep $[h_{N-n}]$
belongs to the $G$ Kronecker product
$\langle\sigma\rangle\times\langle\Sigma_0\rangle$. 
So the problem of finding interactions
that preserve solvability
for part of the states~(\ref{chain}) 
is reduced to carrying out a Kronecker product. 

In what follows we illustrate the above procedure and demonstrate the 
relevance of the PDS notion to nuclear spectroscopy. 
For that purpose, we employ the interacting boson model 
(IBM)~\cite{Iachello87}, 
widely used in the description of low-lying collective states 
in nuclei in terms of $N$ interacting monopole $(s)$ and
quadrupole $(d)$ bosons representing valence nucleon pairs.
The dynamical algebra is $G_{\rm dyn}={\rm U}(6)$
and the symmetry algebra is $G_{\rm sym}={\rm O}(3)$.
The Hamiltonian commutes with the total number operator of $s$- 
and $d$- bosons, $\hat{N} = \hat{n}_s + \hat{n}_d$, 
which is the linear Casimir of U(6). The model accommodates three DS limits 
with leading subalgebras U(5), SU(3), and O(6),
corresponding to typical collective spectra observed in nuclei,
vibrational, rotational, and $\gamma$-unstable, respectively.

\section{\textbf{SU(3)-PDS and axially deformed nuclei}}

The SU(3) DS chain of the IBM 
and related quantum numbers are given by~\cite{Iachello87}
\ba
\begin{array}{ccccc}
{\rm U}(6)&\supset&{\rm SU}(3)&\supset&{\rm O}(3)\\
\downarrow&&\downarrow&&\downarrow\\[0mm]
[N]&&\left (\lambda,\mu\right )& K & L
\end{array} ~.
\label{chainsu3}
\ea 
The multiplicity label $K$ 
is needed for complete classification and corresponds geometrically to the
projection of the angular momentum on the symmetry axis. 
The eigenstates $\vert [N](\lambda,\mu)K,L\rangle$ are obtained with 
a Hamiltonian with SU(3) DS which, for one- and two-body interactions, 
can be transcribed in the form 
\ba
\hat{H}_{\rm DS} &=& h_{2}\left [-\hat C_{2}(\rm SU(3)) 
+ 2\hat N (2\hat N+3)\right ]
+ C\, \hat C_{2}(\rm O(3)) ~.
\label{hDSsu3}
\ea
Here $\hat{C}_{2}(G)$ denotes the quadratic Casimir operator of $G$. 
The spectrum of $\hat{H}_{\rm DS}$ is completely solvable with eigenenergies
\ba
E_{\rm DS} &=& 
h_{2}\,\left [ -(\lambda^2 +\mu^2 +\lambda\mu 
+ 3\lambda + 3\mu) + 2N(2N+3)\right ] + CL(L+1) ~.
\label{eDSsu3}
\ea 
The spectrum resembles that of an axially-deformed rotor 
and the corresponding eigenstates are arranged in SU(3) multiplets. 
In a given SU(3) irrep $(\lambda,\mu)$, each $K$-value is associated 
with a rotational band and states 
with the same L, in different $K$-bands, are degenerate. 
In particular, the lowest SU(3) irrep $(2N,0)$, 
contains the ground band $g(K=0)$ and the irrep 
$(2N-4,2)$ contains degenerate $\beta(K=0)$ and $\gamma(K=2)$ bands.
This $K$-band degeneracy is a characteristic feature of the SU(3) 
limit of the IBM which, however, is not commonly observed 
in deformed nuclei. 
In the IBM framework, with at most two-body interactions, one
is therefore compelled to break the SU(3) DS. 
\begin{figure}[t]
\includegraphics[height=6cm]{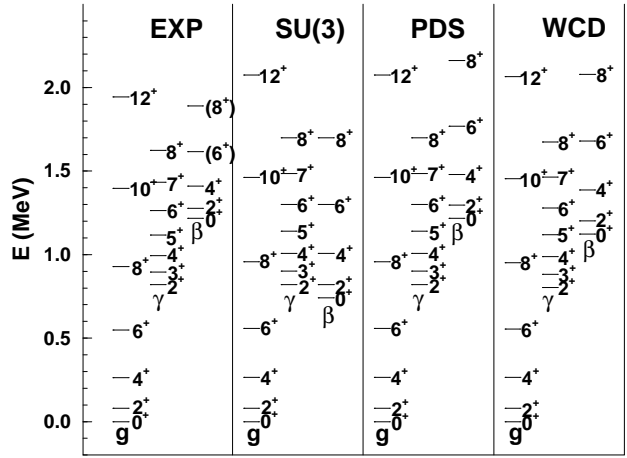}
\caption{
\small
Spectra of $^{168}$Er. Experimental energies
(EXP) are compared with IBM calculations in an exact SU(3) dynamical 
symmetry [SU(3)], in a broken SU(3) symmetry (WCD) 
and in a partial dynamical SU(3) symmetry (PDS). 
The latter employs the Hamiltonian of Eq.~(\ref{hPDSsu3}), 
with $N=16$ and $h_2=4,\,\eta_0=4,\,C=13$ keV~\cite{lev96}. 
\label{fig1er168}}
\end{figure}

To secure solvability and good SU(3) symmetry for the ground band, 
we follow the PDS algorithm and identify 
$n$-boson operators which annihilate all $L$-states in the 
SU(3) irrep $(2N,0)$. For that purpose, we 
consider the following two-boson SU(3) tensors, 
$B^{\dagger}_{[n](\lambda,\mu)\kappa;\ell m}$, with $n=2$, 
$(\lambda,\mu)=(0,2)$ and 
angular momentum $\ell =0,\,2$
\ba
B^{\dagger}_{[2](0,2)0;00} \propto 
P^{\dagger}_{0} = d^{\dagger}\cdot d^{\dagger} - 2(s^{\dagger})^2
\;\; , \;\;\; 
B^{\dagger}_{[2](0,2)0;2m} \propto 
P^{\dagger}_{2m} = 2d^{\dagger}_{m}s^{\dagger} + 
\sqrt{7}\, (d^{\dagger}\,d^{\dagger})^{(2)}_{m} ~.
\;
\ea
\label{PL}
The corresponding annihilation operators,  $P_0$ and $P_{2m}$, satisfy
\ba
P_{0}\,\vert [N](2N,0)K=0, L\rangle &=& 0~,
\nonumber\\
P_{2m}\,\vert [N](2N,0)K=0, L\rangle &=& 0~, 
\;\;\;\;\;
L=0,2,4,\ldots, 2N~.
\qquad 
\label{P0P2}
\ea
In addition,  $P_0$ satisfies 
\ba
P_{0}\,\vert [N](2N-4k,2k)K=2k, L\rangle = 0 ~,\;\;\;\;
L=K,K+1,\ldots, (2N-2k)~.
\label{P0}
\ea
For $k> 0$ the indicated $L$-states 
span only part of the SU(3) irreps 
$(\lambda,\mu)=(2N-4k,2k)$ and form the rotational members of excited 
$\gamma^{k}(K=2k)$ bands. The combination 
$P^{\dagger}_{0}P_{0} + P^{\dagger}_{2}\cdot \tilde{P}_{2}$ 
is completely solvable in SU(3) and is simply the first ($h_2$) term 
in $\hat{H}_{\rm DS}$, Eq.~(\ref{hDSsu3}), related to 
$\hat C_{2}(\rm SU(3))$. 
The SU(3)-PDS Hamiltonian is thus given by
\ba
\hat{H}_{\rm PDS} &=& 
\hat{H}_{\rm DS} + \eta_0\, P^{\dagger}_{0}P_{0} ~.
\label{hPDSsu3}
\ea
The relations in Eqs.~(\ref{P0P2})-(\ref{P0}) ensure that 
$\hat{H}_{\rm PDS}$ retains solvable ground $g(K=0)$ and $\gamma^{k}(K=2k)$ 
bands with good SU(3) symmetry $(2N-4k,2k)$ and 
energies as in Eq.~(\ref{eDSsu3}). 
The remaining eigenstates of $\hat{H}_{\rm PDS}$, including the 
$\beta(K=0)$ band, are mixed. 

The SU(3)-PDS spectrum of the Hamiltonian~(\ref{hPDSsu3}) is compared 
with the empirical spectrum of $^{168}$Er in Fig.~\ref{fig1er168}. 
As shown, the undesired $\beta$-$\gamma$ degeneracy is lifted and the 
PDS fit is of comparable quality to that of a broken-SU(3) calculation. 
Since the wave functions of the solvable 
states~(\ref{P0P2})-(\ref{P0}) are known, one can obtain 
{\it analytic} expressions for matrix elements of observables between them. 
In particular, the resulting B(E2) values are found to be in excellent 
agreement with experiment~\cite{lev96}, thus 
confirming the relevance of SU(3)-PDS to the spectroscopy 
of $^{168}$Er.
\begin{figure}[t]
\hspace{-0.13cm}
\includegraphics[width=0.9\textwidth]{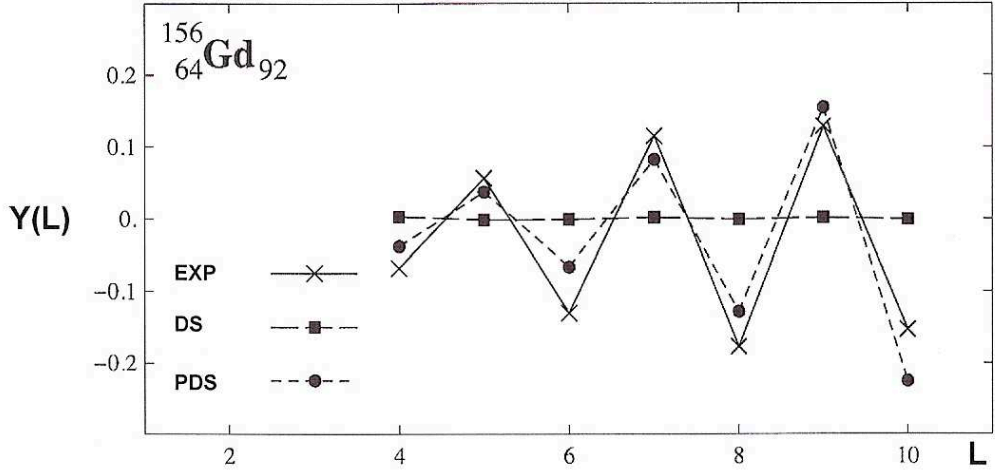}
\caption{Experimental (EXP) odd-even staggering in the $\gamma$-band 
of $^{156}$Gd, compared with SU(3)-DS and SU(3)-PDS calculations. 
The latter employs the Hamiltonian of Eq.~(\ref{hPDSsu3-2}), with 
$N=12$ and $h_2=7.6,\, C= 10.46,\, \eta_2=-0.24,\, \eta_3=1.68$ 
keV~\cite{LevRamVan12}. The staggering index $Y(L)$ is defined in 
the text. 
}
\end{figure}

The dynamical symmetry expression, Eq.~(\ref{eDSsu3}), implies 
a pure rotor spectrum with characteristic $L(L+1)$ in-band splitting. 
Such a pattern is observed in the empirical spectrum of the ground 
and $\beta$ bands in $^{156}$Gd, but the $\gamma$ band exhibits 
considerable odd-even staggering (OES). 
As shown in Fig.~3, the empirical staggering index, 
$Y(L) = \frac{(2L-1)}{L}\,\frac {\left [\,E(L)-E(L-1)\,\right ]}
{\left [\,E(L)-E(L-2)\,\right ]} -1$, displays a pronounced zigzag pattern, 
in marked deviation from a pure rotor for which $Y(L)=0$.
The fact that the SU(3)-DS is obeyed only in selected bands, 
highlights its partial nature. 
Following the PDS algorithm, 
we look for $n$-boson operators which annihilate the states in the 
$g(K=0)$ and $\beta(K=0)$ bands. 
For $n=3$, we identify the following 
SU(3) tensors, $\hat{B}^{\dagger}_{[n](\lambda,\mu)\kappa; \ell m}$, 
with $(\lambda,\mu)=(2,2)$ and $\ell=2,\,3$ 
\ba
\hat B^\dag_{[3](2,2)2;\ell m} 
\propto 
W^{\dag}_{\ell m} = (P^{\dag}_{2}d^{\dag})^{(\ell)}_{m}\;\;\;\;\; 
\ell=2,\,3 \qquad
\ea
which satisfy
\ba
&&W_{\ell m}\vert [N](2N,0)K=0,L\rangle = 0
\qquad\;\;\;
L=0,2,4,\ldots, 2N
\qquad 
\nonumber\\[1mm]
&&W_{\ell m}\vert [N](2N-4,2)K=0,L\rangle = 0 
\;\;\;\;\;
L=0,2,4,\ldots, (2N-4)~.
\qquad
\label{W2W3}
\ea
The Hamiltonian with SU(3)-PDS, 
now involving three-body terms, reads~\cite{LevRamVan12}
\ba
\hat{H}_{\rm PDS} &=& \hat{H}_{\rm DS}
+\eta_{2}\,W^{\dag}_{2}\cdot \tilde{W}_{2} 
+\eta_{3}\,W^{\dag}_{3}\cdot \tilde{W}_{3} ~.
\label{hPDSsu3-2}
\ea
The cubic Casimir operator of SU(3), $\hat{C}_{3}(\rm SU(3))$, 
can also be included in $\hat{H}_{\rm DS}$. 
The relations in Eq.~(\ref{W2W3})
ensure that 
$\hat{H}_{\rm PDS}$ retains solvable ground $g(K=0)$ and $\beta(K=0)$ 
bands with good SU(3) symmetry $(2N,0)$ and $(2N-4,2)$, respectively, 
and energies as in Eq.~(\ref{eDSsu3}). 
Other eigenstates, including members of the $\gamma$ band, are mixed. 
As seen in Fig.~3, the SU(3)-PDS calculation 
can adequately reproduce the observed OES in $^{156}$Gd~\cite{LevRamVan12}. 
In this case, the staggering arises from the coupling of the $\gamma$ band 
with higher excited bands. Other approaches advocating the coupling of the 
$\gamma$ band to the $\beta$ band~\cite{Bona88} or to the ground 
band~\cite{Minkov00}, 
cannot describe the OES in this nucleus. 

\section{\textbf{O(6)-PDS and ${\rm \gamma}$-unstable nuclei}}
\label{subsec:o6PDS}

The O(6) DS chain of the IBM and related quantum numbers are given 
by~\cite{Iachello87}
\ba
\begin{array}{ccccccc}
{\rm U}(6)&\supset&{\rm O}(6)&\supset&{\rm O}(5)&
\supset&{\rm O}(3)\\
\downarrow&&\downarrow&&\downarrow&&\downarrow\\[0mm]
[N]&&\langle\Sigma\rangle&&(\tau)&n_\Delta& L
\end{array} ~.
\label{chaino6}
\ea
The multiplicity label $n_{\Delta}$ is needed for complete 
classification. A completely solvable spectrum with eigenstates 
$\vert [N]\langle\Sigma\rangle(\tau)n_\Delta L\rangle$
and eigenenergies $E_{\rm DS}$, is obtained with a Hamiltonian
with O(6) DS, which has the form 
\ba
\hat{H}_{\rm DS} &=& h_{0}\left [-\hat C_{2}({\rm O(6)}) 
+ \hat N (\hat{N} +4)\right ]
+ B\, \hat{C}_{2}({\rm O(5)}) + C\,\hat{C}_{2}({\rm O(3)}) ~,
\nonumber\\[1mm] 
E_{\rm DS} &=& 
h_{0}\,\left [-\Sigma(\Sigma+4) + N(N+4)\right ]
+ B\,\tau(\tau+3) +\, C\,L(L+1) ~.
\label{hDSo6}
\ea
The spectrum resembles that of a $\gamma$-unstable deformed rotor, 
where the states are arranged in bands with O(6) quantum number 
$\Sigma=N-2v$, $(v=0,1,2,\ldots)$. 
The ground band ($v=0$) corresponds to the O(6) irrep with $\Sigma=N$. 
The O(5) and O(3) terms in $\hat{H}_{\rm DS}$~(\ref{hDSo6}), 
govern the in-band rotational splitting. 

\begin{figure*}[t]
\leavevmode
\includegraphics[width=0.93\linewidth]{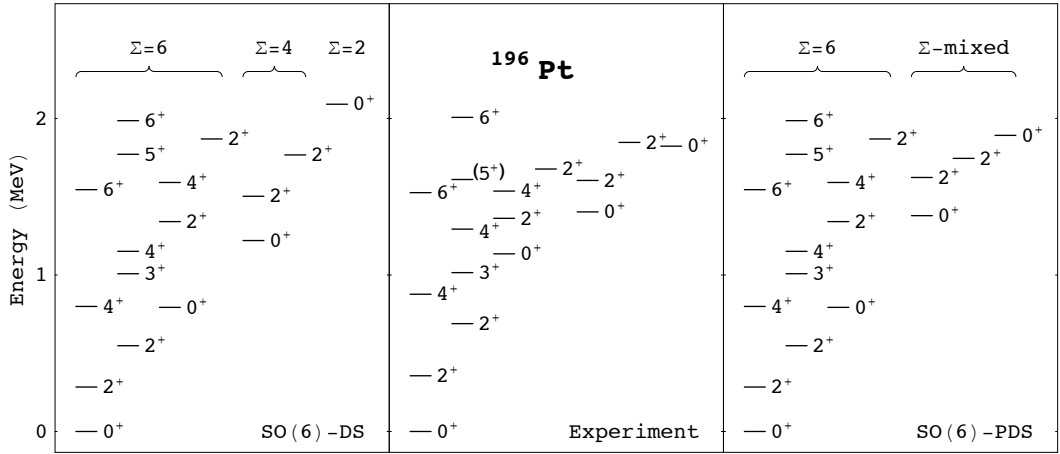}
\caption{
\small
Observed spectrum of $^{196}$Pt compared with the calculated spectra 
of $\hat H_{\rm DS}$~(\ref{hDSo6}), 
with O(6) dynamical symmetry (DS), 
and of $\hat H_{\rm PDS}$~(\ref{hPDSo63bod}) with 
O(6) partial dynamical symmetry (PDS). 
The parameters in $\hat H_{\rm DS}$ $(\hat H_{\rm PDS})$ are
$h_0=43.6\, (30.7)$, $B=44.0\, (44.0)$, $C=17.9\, (17.9)$, 
and $\rho_{0}=0\, (8.7)$ keV. The boson number is $N=6$ 
and $\Sigma$ is an O(6) label~\cite{RamLevVan09}.}
\label{fig2pt196}
\end{figure*}

The O(6)-DS limit provides a good description of the empirical spectrum 
and E2 rates in $^{196}$Pt, for states in the ground band $(\Sigma=N)$.
This observation was the basis of the claim~\cite{Cizewski78} 
that the O(6)-DS is manifested empirically in this nucleus. 
However, as shown in Fig.~\ref{fig2pt196}, 
the resulting fit to energies of excited bands is quite poor.
The $0^+_1$, $0^+_3$, and $0^+_4$ levels of $^{196}$Pt
at excitation energies 0, 1403, 1823 keV, respectively,
are identified as the bandhead states
of the ground $(v=0)$, first- $(v=1)$
and second- $(v=2)$ excited vibrational bands~\cite{Cizewski78}.
Their empirical anharmonicity,
defined by the ratio $R=E(v=2)/E(v=1)-2$,
is found to be $R=-0.70$.
In the O(6)-DS limit these bandhead states
have $\tau=L=0$ and $\Sigma=N,N-2,N-4$, respectively.
The anharmonicity $R=-2/(N+1)$,
as calculated from Eq.~(\ref{hDSo6}), is fixed by $N$.
For $N=6$, which is the appropriate boson number for $^{196}$Pt,
the O(6)-DS value is $R=-0.29$,
which is in marked disagreement with the empirical value. 
Large anharmonicities can be incorporated in the IBM 
only by the inclusion of at least cubic terms in the
Hamiltonian~\cite{ramos00b}. One is therefore confronted with the 
need to select suitable higher-order terms
that can break the DS in excited bands but preserve it in the ground band. 
These are precisely the defining 
properties of a PDS. Following the general algorithm, 
we look for $n$-boson operators which annihilate all 
the states in the O(6) ground band. 
For $n=3$, there are two such O(6) tensors, 
$\hat B^\dag_{[n]\langle\sigma\rangle(\tau)n_{\Delta};\ell m}$, 
with $\sigma=1$, $(\tau,\ell)=(0,0)$ and $(\tau,\ell)=(1,2)$, given by 
\ba
\hat B^\dag_{[3]\langle1\rangle(0)0;00}
\propto P^{\dag}_{0}s^\dag 
\;\;\;\; , \;\;\;\;
\hat B^\dag_{[3]\langle1\rangle(1)0;2m}
\propto P^{\dag}_{0}d^\dag_m ~.
\ea
Here $P^{\dagger}_{0} = d^{\dagger}\cdot d^{\dagger} - (s^{\dagger})^2 ~$, 
is an O(6) scalar and $P^{\dag}_{0}P_0$ is simply the first ($h_0$) term 
in $\hat{H}_{\rm DS}$, Eq.~(\ref{hDSo6}), 
related to $\hat{C}_{2}({\rm O(6)})$. 
The operators, $sP_{0}$ and $d_{m}P_0$,  
annihilate all $(\tau,L)$-states in the O(6) irrep  
$\langle\Sigma\rangle= \langle N\rangle$. This 
is ensured by the fact that
\ba
P_{0}\,\vert [N]\langle N \rangle (\tau)n_{\Delta}L\rangle = 0 ~, 
\;\;\;\;\; \tau=0,1,2,\ldots,N~.
\label{P0o6}
\ea
The only three-body interactions that are partially solvable in O(6)
are thus $P^{\dag}_{0}\hat n_s P_{0}$
and $P^{\dag}_{0}\hat n_d P_{0}$. 
Since the combination $P^{\dag}_{0}(\hat n_s+\hat n_d)P_{0}
= (\hat{N} -2)P^{\dag}_{0}P_{0}$
is completely solvable in O(6), we can transcribe the O(6)-PDS 
Hamiltonian in the form
\ba
\hat{H}_{\rm PDS}=\hat{H}_{\rm DS} + 
\rho_{0}\,P^{\dag}_{0}\hat{n}_s P_{0} ~.
\label{hPDSo63bod}
\ea

The spectrum of $\hat{H}_{\rm PDS}$~(\ref{hPDSo63bod}) 
is shown in Fig.~\ref{fig2pt196}.
The states belonging to the $\Sigma=N=6$ multiplet remain solvable
with energies which obey the same DS expression, 
Eq.~(\ref{hDSo6}). States with $\Sigma < 6$ are generally admixed 
but agree better with the data than in the DS calculation. 
For example, the bandhead states of the first- (second-) excited bands 
have the O(6) decomposition 
$\Sigma=4$: $76.5\%\,(19.6\%)$, 
$\Sigma=2$: $16.1\%\,(18.4\%)$, 
and $\Sigma=0$: $7.4\%\,(62.0\%)$. 
Thus, although the ground band is pure, 
the excited bands exhibit strong O(6) breaking. 
The calculated O(6)-PDS anharmonicity for these bands is $R=-0.63$, 
much closer to the empirical value, $R=-0.70$. 
It should be emphasized that not only the energies 
but also the wave functions of the $\Sigma=N$ states remain unchanged
when the Hamiltonian is generalized from DS to PDS.
Consequently, the E2 rates for transitions among this class of states
are the same in the DS and PDS calculations~\cite{RamLevVan09}. 
Thus, the additional three-body ($\rho_0$) term in the 
Hamiltonian~(\ref{hPDSo63bod}), does not spoil the good O(6)-DS 
description for this segment of the spectrum. 

\section{\textbf{Summary and Conclusions}}

The notion of partial dynamical symmetry (PDS) extends and complements 
the fundamental concepts of exact and dynamical symmetries. 
It addresses situations in which a prescribed symmetry is neither 
exact nor completely broken. When found, such intermediate symmetry structure 
can provide analytic solutions and quantum numbers for a portion of the 
spectra, thus offering considerable insight into complex dynamics.

On phenomenological grounds, having at hand a concrete algorithm for 
identifying and constructing Hamiltonians with PDS, is a valuable asset. 
It provides selection criteria for the a priori huge number of 
possible symmetry-breaking terms, accompanied by a rapid proliferation 
of free-parameters. This is particularly important in complicated 
environments when many degrees of freedom take part in the dynamics 
and upon inclusion of higher-order terms in the Hamiltonian. 
In the IBM examples considered, there are 17 possible three-body 
interactions, yet only a few terms satisfy the PDS requirements.
Futhermore, Hamiltonians with PDS break the dynamical symmetry (DS) 
but retain selected solvable eigenstates with good symmetry. The 
advantage of using interactions with a PDS is that they can be introduced, 
in a controlled manner, without destroying results previously obtained 
with a DS for a segment of the spectrum. These virtues 
greatly enhance the scope of applications of algebraic modeling 
of quantum many-body systems.
On a more fundamental level, PDS can offer a possible clue to 
the deep question of how simple features emerge from complicated dynamics.

PDSs are not restricted to a specific model but can be applied 
to any quantal systems of interacting particles, bosons, as demonstrated 
in the present contribution, 
and fermions~\cite{escher00,rowe01,isa08}. They are also relevant 
to quantum phase transitions~\cite{lev07} 
and to the study of mixed systems with 
coexisting regularity and chaos~\cite{walev93,levwhe96}.

\section{\textbf{Acknowledgments}}
It is a pleasure and honor to dedicate this contribution to F. Iachello 
on the occasion of his 70th birthday. 
Franco's innovative approach to physics, emphasizing the 
deep connection between symmetries and beauty, as manifested in theory 
and experiment, has influenced and inspired the ideas discussed here. 
Segments of the reported results were obtained in collaboration with 
J.~E.~Garc\'\i a-Ramos (Huelva) and P.~Van~Isacker~(GANIL). 
This work is supported by the Israel Science Foundation.

\end{document}